\documentclass[aps,pre,10pt,tightenlines,superscriptaddress,amsmath,amssymb,final,letterpaper,notitlepage,nofootinbib]{revtex4-1}
\usepackage{graphicx}
\usepackage{subfigure}
\usepackage{epstopdf}
\usepackage{times}
\usepackage{bm}
\usepackage[dvipsnames]{xcolor}
\usepackage{tcolorbox}
\usepackage{hyperref}
\usepackage{mathtools}
\usepackage{comment}
\usepackage{amsmath}
\usepackage[pagewise]{lineno}
\usepackage{placeins}

\graphicspath{{Figures/}}

\begin{document}

\title{Immunity-induced criticality of the genotype network of influenza A (H3N2) hemagglutinin}

\author{Blake J.M. Williams}
\affiliation{Vermont Complex Systems Center, University of Vermont, Burlington VT, USA}
\author{C. Brandon Ogbunugafor}
\affiliation{Vermont Complex Systems Center, University of Vermont, Burlington VT, USA}
\affiliation{Department of Ecology and Evolutionary Biology, Yale University, New Haven CT, USA}
\author{Benjamin M. Althouse}
\affiliation{Institute for Disease Modeling, Global Health, Bill \& Melinda Gates Foundation, Seattle WA, USA}
\affiliation{University of Washington, Seattle, WA, USA}
\affiliation{New Mexico State University, Las Cruces, NM, USA}
\author{Laurent H\'ebert-Dufresne}%
\affiliation{Vermont Complex Systems Center, University of Vermont, Burlington VT, USA}
\affiliation{Department of Computer Science, University of Vermont, Burlington VT, USA}

\begin{abstract}
Seasonal influenza kills hundreds of thousands every year, with multiple constantly-changing strains in circulation at any given time. A high mutation rate enables the influenza virus to evade recognition by the human immune system, including immunity acquired through past infection and vaccination. Here, we capture the genetic similarity of influenza strains and their evolutionary dynamics with genotype networks. We show that the genotype networks of influenza A (H3N2) hemagglutinin are characterized by heavy-tailed distributions of module sizes and connectivity, suggesting critical-like behavior. We argue that: (i) genotype networks are driven by mutation and host immunity to explore a subspace of networks predictable in structure, and (ii) genotype networks provide an underlying structure necessary to capture the rich dynamics of multistrain epidemic models. In particular, inclusion of strain-transcending immunity in epidemic models is dependent upon the structure of an underlying genotype network. This interplay suggests a self-organized criticality where the epidemic dynamics of influenza locates critical-like regions of its genotype network. We conclude that this interplay between disease dynamics and network structure might be key for future network analysis of pathogen evolution and realistic multistrain epidemic models.
\end{abstract}

\maketitle

\section{Introduction}

Each year, seasonal influenza results in  290,000 to 650,000 deaths globally, 9 million to 36 million cases in the United States alone, and results in significant economic burdens~\cite{Putri:2018,Molinari:2007,Rolfes:2018}. Despite widespread vaccination and increased surveillance efforts in recent years, influenza continues to show prominent seasonality in temperate regions and causes a year-round burden in tropical regions~\cite{Iuliano:2018, Nair:2011}. 

Influenza viruses (INFV) mutate rapidly with antigenic drifts and shifts, leading to the frequent emergence of new strains that are different enough to escape recognition by host immunity~\cite{Guan2010}. As a result, we see frequent epidemics and necessitate yearly updates to vaccine strains based on sequencing data and future projections~\cite{ barr2010epidemiological, klimov2012recommendations, barr2014recommendations}. Optimal vaccine strain selection is dependent upon the ability to both forecast prevalent future strains and select a limited number of vaccine strains, such that these strains offer optimal immune protection by leveraging strain-transcending immunity~\cite{Carrat2007,Hensley2014}. Modern seasonal INFV vaccines induce antibodies for three to four unique strains of INFV, providing direct immunity for these strains and some cross-protective (or strain-transcending) effects towards antigenically similar strains. Similarly, these antibodies are induced in response to a clinical influenza infection~\cite{Peeters:2017}. 

INFV epidemiology has benefited from decades of research using phylogenetics and molecular evolution to carefully interrogate features of INFV evolution ~\cite{Taubenberger:2010, webster1992evolution, Nelson:2007}. Exercises in applied evolutionary theory have served as validations for the use of molecular methods towards meaningful predictive evolution~\cite{lassig2017predicting, morris2018predictive}. These methods, in combination with larger data sets, offer increasingly accurate probabilistic models for INFV evolution. As effective as they have been, these approaches are based on particular population genetic assumptions and limitations. For example, tree-based methods are necessarily acyclic and as such do not fully capture the relatedness of strains.

Phylodynamic approaches have features of neutral networks, defined by genotypes that evolve via drifting through epochal evolution~\cite{koelle2006epochal, van2006influenza}.  Genotype networks constitute another approach used to study INFV evolution, and are built on different assumptions and constraints than other approaches ~\cite{Wagner:2014,Luksa2014,Neher:2016,Smith:2004, fonville2014antibody}. Previous networks have been constructed from the highly antigenic hemagglutinin (HA) protein sequences of INFV~\cite{Wagner:2014}. The networks revealed features not well represented in phylogenetic trees, such as identical trait evolution in separate lineages (convergent evolution). Their many strengths aside, these networks were prone to fragmentation in the presence of low sampling rates, reducing the number of observed plausible evolutionary pathways. Sampling has increased dramatically in the last decade, which now allows for a more accurate account of the evolution of INFV genotype networks. 

In this study, we utilize a large modern data set of INFV H3N2 sequences (over 28,000) and a genotype network approach to capture the genetic relationship between the 2010 and 2020 INFV H3N2 strains and their evolutionary dynamics. Sequences of the highly antigenic HA protein of INFV A (H3N2) are used to analyze the structure and temporal evolution of the genotype network and its exploration of genotype space. Finally, a multi-strain epidemic model is implemented to explore how the density and distribution of edges (or mutation pathways) determines epidemic potential in the context of strain-transcending immunity. We demonstrate the existence of a fundamental structure underlying INFV genotype space, one that captures temporal features of virus evolution and suggests underlying predictability. In doing so, we fortify the relevance of genotype networks as a meaningful approach to the study of virus evolution, one that can complement mathematical and phylodynamic approaches in future efforts to study and predict the dynamics of evolution of INFV and other RNA viruses.

\section{Data and model}

\subsection{Network generation}
Protein sequences were obtained for complete INFV A (H3N2) HA samples from the Influenza Research Database~\cite{fludb}. Samples acquired from the Influenza Research Database are sourced from databases that include NCBI GenBank and RefSeq. Samples were obtained on January 16 2020 and restricted to a collection date of January 4 1999 through October 1 2019 and collected from human hosts only. A 3 month delay between final sample collection date and data retrieval date was implemented to account for delays in data reporting. 

A total of 30,175 sequenced samples for HA were obtained. Sequences were further restricted to allow for the precise genetic sequence comparison required for network edge construction. Samples with  missing or uncertain residues ($n= 1,278$) and sequences with more or less than 566 amino acids ($n=17$) were removed. The remaining 28,880 samples were condensed into set $V$ of 6,494 unique sequences.

 The number of differing amino acids across all sites for sequences $v$ and $w$, $d_{v,w}$, was found for all pairs of sequences of length $l=566$: 
 $$  d_{v,w}=  \sum_{i=1}^{l}{x}, \text{where } x= \begin{cases}
1, & \mbox{if } v_i \neq w_i \\
0, & \mbox{if } v_i = w_i
\end{cases}  \ \ \ \ v,w \in V  $$ 
An edge $e_{v,w}$ is formed if $d_{v,w}=1$. Each edge indicates a plausible, but not definitive, mutation pathway between two viable strains that requires one point mutation, thus no intermediate strains nor multi-mutation events. The resulting genotype network is defined as $G=(V,E)$, where $E$ is the set of all edges $e_{v,w}$. 

Temporal analyses restricted data by year using seasonal trends of the Northern Hemisphere, given its dominance of the data set. Sequences were binned according to a 5-year window, where each year consisted of July 1 through June 30 of the following year. For example, a 5-year window centered on 2010 would contain sequences from July 2007 through June 2012.

\subsection{Multistrain epidemic model}



We assume that the epidemiological dynamics of INFV follow the classic Susceptible-Infectious-Recovered-Susceptible (SIRS) model, but we introduce an underlying genotype network which defines potential mutations and allows strain-transcending immunity. An individual infected with strain $i \in [1,N]$ can cause a mutation at a rate $\mu$ to a strain $j\in \mathcal{N}_i$, where $\mathcal{N}_i$ is the set of first network neighbors of strain $i$. 

All strains spread concurrently in a well-mixed host population. Individuals are susceptible ($S$) if they possess no previous immunity. Each susceptible individual progresses to infectious state $I_i$, corresponding to strain $i$, at a rate $\beta I_i$. The basic transmission rate $\beta$ is held constant for all strains, as we focus on neutral evolution (antigenic drift) as a first approximation. 

Infectious individuals in $I_i$ will either: (i) recover at rate $\gamma$ to state $R_i$ and acquire full immunity for strain $i$ and partial immunity to other strains $j\neq i$; or (ii) undergo a mutation to strain $j$ at a rate $\mu$ for all strains $j$ in $\mathcal{N}_i$. Recovered individuals in $R_i$ will either: (i) lose immunity and progress back to $S$ at rate $\alpha$, or (ii) get infected with strain $j\neq i$ and progress to $I_j$ at a reduced rate $\beta^*$ due to their partial immunity. Specifically, $\beta^*$ is an exponentially decaying function of genetic distance between strains $i$ and $j$,
\begin{equation}
\beta^*_{ij} = \beta \left( 1-e^{-x_{ij}/\Delta}  \right) \nonumber
\end{equation}
where $x_{ij}$ is the network distance between strain $i,j$ (shortest path between strains $i$ and $j$ in the genotype network, different from $d_{v,w}$ used above) and $\Delta$ is the characteristic length of immunity ($0<\Delta<\infty$) as it transcends specific strains over the genotype network. Note that we make the assumption that an individual's immune response is set by the most recent infection as accounting for a full immune history would result in $N!$ possible immune states. 

The model assumes that: (i) an individual may be infected by at most one strain at a time, (ii) an individual's immune response is determined by the strain responsible for their last infection, and (iii) transcendence of immunity decays exponentially as a function of the distance between strains. The model was implemented with a system of differential equations containing one susceptible state and an infected and recovered state for each strain. The dynamical system describing this model are presented in Materials and Methods, and its dynamics were studied in Ref.~\cite{williams2021localization}. 

The model itself can run over any genotype network defined as a number of strains $i \; \in \; [1,N]$ and a set of neighboring strains $j \; \in \; \mathcal{N}_i$ for each of strain. In what follows, we therefore couple the model with known generative models of networks that can help explain some key network features found in the genotype data.

\begin{figure*}[t!]
\centering
\includegraphics[width = 0.9\linewidth]{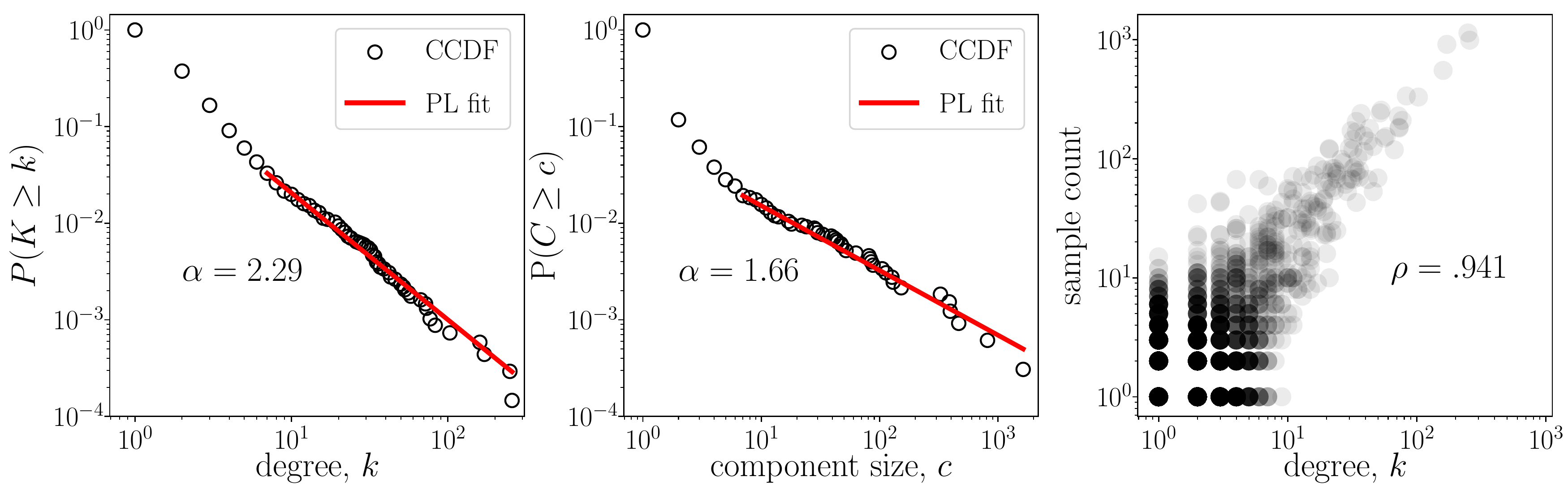}
\caption{\textbf{INFV A (H3N2) HA genotype network degree and component size distribution.}
\textbf{(left)} Complementary cumulative distribution function (CCDF) of degrees. The tail of degree distribution does not significantly differ from a power-law distribution with $\alpha_k=2.29$ for $k_{\textrm{min}}=7$ ($p=0.11$, $\alpha_{\textrm{significance}}=0.05$, $10^3$ repetition Kolmogorov-Smirnov test).  \textbf{(center)} CCDF of component sizes. Component size distribution does not significantly differ from a power-law distribution with $\alpha_c=1.66$ for $c_{\textrm{min}}=7$ ($p=0.59$, $\alpha_{\textrm{significance}}=0.05$, $10^3$ repetition Kolmogorov-Smirnov test). \textbf{(right)} Sample count of a sequence vs. degree $k$ of corresponding node. Sample count is highly correlated with node degree $(r=0.941)$. }
\label{fig:features}
\end{figure*}

\section{Results}

\subsection{INFV A (H3N2) HA genotype network}

The INFV A (H3N2) HA genotype network represents 28,880 samples of HA, resulting in 9,714 nodes (unique strains), 7,599 edges (possible point mutations between strains), and 3,262 connected components, of which 384 consist of more than one node. With $29.6\%$ of nodes of degree $k=0$ and $44.0\%$ of $k=1$, the network features a skewed degree distribution, stretching up to a maximum degree of $k=256$. The tail of the complementary cumulative distribution function (CCDF) of degree, $P(K\geq k)$, exhibits power-law behavior: $P(K\geq k) \; \propto k^{-\alpha_k}$ with an estimated scale exponent $\alpha_k=2.29$, Fig. \ref{fig:features}(left panel). This is in agreement with the heavy-tailed degree distribution found by Wagner in the largest connected component of a smaller data set from 2002-2007~\cite{Wagner:2014}. This degree distribution suggests that in a generative model of such a network, approximately linear preferential attachment could underlie the growth of the genotype network~\cite{BA,hebert2016constrained}. Linear attachment is a critical mechanism such that a growing network may produce power-law degree distributions, versus exponential distributions under sub-linear attachment or even condensating in a star graph for super-linear attachment~\cite{krapivsky2001degree, krapivsky2001organization}.	

The distribution of component sizes of the genotype network is similarly skewed. The tail of the CCDF of component sizes $P(C\geq c)$ follows a power-law distribution, where $P(C\geq c) \propto c^{-\alpha_c}$ with scale exponent $\alpha_c=1.66$, Fig.~\ref{fig:features}(center panel). This scaling may also be suggestive of another self-organized critical process in the formation of the genotype network.

The degree of a node and the number of times its corresponding sequence was sampled are highly correlated, Fig.~\ref{fig:features}(right panel). This suggests that high degree nodes are robust to reduced sampling, given that the duplicate sample count of a strain may be a proxy for its population prevalence. This is akin to the robustness of scale-free networks towards random failures, with sample count taking place of degree~\cite{robust:2000, albert2000error}. Consequently, increased surveillance has likely identified more strains of low prevalence given that low-degree strains are more likely to be removed from the network with random sub-sampling.  

The network contains numerous cycles amidst its tree-like structure. Its 500 triangles indicate mutations at the same site between 3 sequences, while sparse squares indicate potential convergent evolution~\cite{Wagner:2014}. These structures are clearly displayed in genotype networks, while phylogenetic tree construction do not include convergent evolution structurally. The treelike topology of the network prevents longer cycles from forming. Further network summary statistics are shown in Table 1 for the enture network $G$ and the giant component $GC$. The triangles are captured by global clustering $C_{global}$, which is equivalent to the proportion of triplets (3 connected nodes) that form a closed triangle. 

  \begin{table}[h!]
  \begin{center}
\begin{tabular}{c c c c c c c c c}
\hline
  & $n$ & $m$ & $\langle k \rangle$ & $k_{\textrm{max}}$ & $D$ & $C_{\textrm{Global}}$  & $r$ & \\ \hline
  \textbf{\textit{G}}& 9714 & 7599 & 1.86 & 257 & - & 0.0096 & -0.13 \\
  \textbf{\textit{GC}}&  1629 &2225  &2.73  & 257 & 17  &0.0010 & -0.20 \\
  
\end{tabular}
\end{center}
\caption[Genotype network statistics]{Statistics of the entire network \textit{G} and its giant component \textit{GC}: Number of nodes $n$ and edges $m$ as well as average degree $\langle k \rangle$, maximum degree $k_{\textrm{max}}$, diameter $D$, clustering coefficient $C_{\textrm{Global}}$ and degree correlations (assortativity coefficient) $r$.}
\label{net_stats}
\end{table}

The degree assortativity $r$ represents the correlation between the degree of a node and that of its neighbors (Table 1). A negative value for both the entire network $G$ and the giant component $GC$ indicate that high degree nodes tend to attach to low degree nodes. Future investigation of vaccination strategies may consider potential high-degree hubs as immunization targets, given their high neighbor count and the transcendence of immunity in INFV A (H3N2) viruses, as well as potential bridges between highly connected regions of the network.

\begin{figure}[t!]
\centering
\includegraphics[width = 0.8\linewidth]{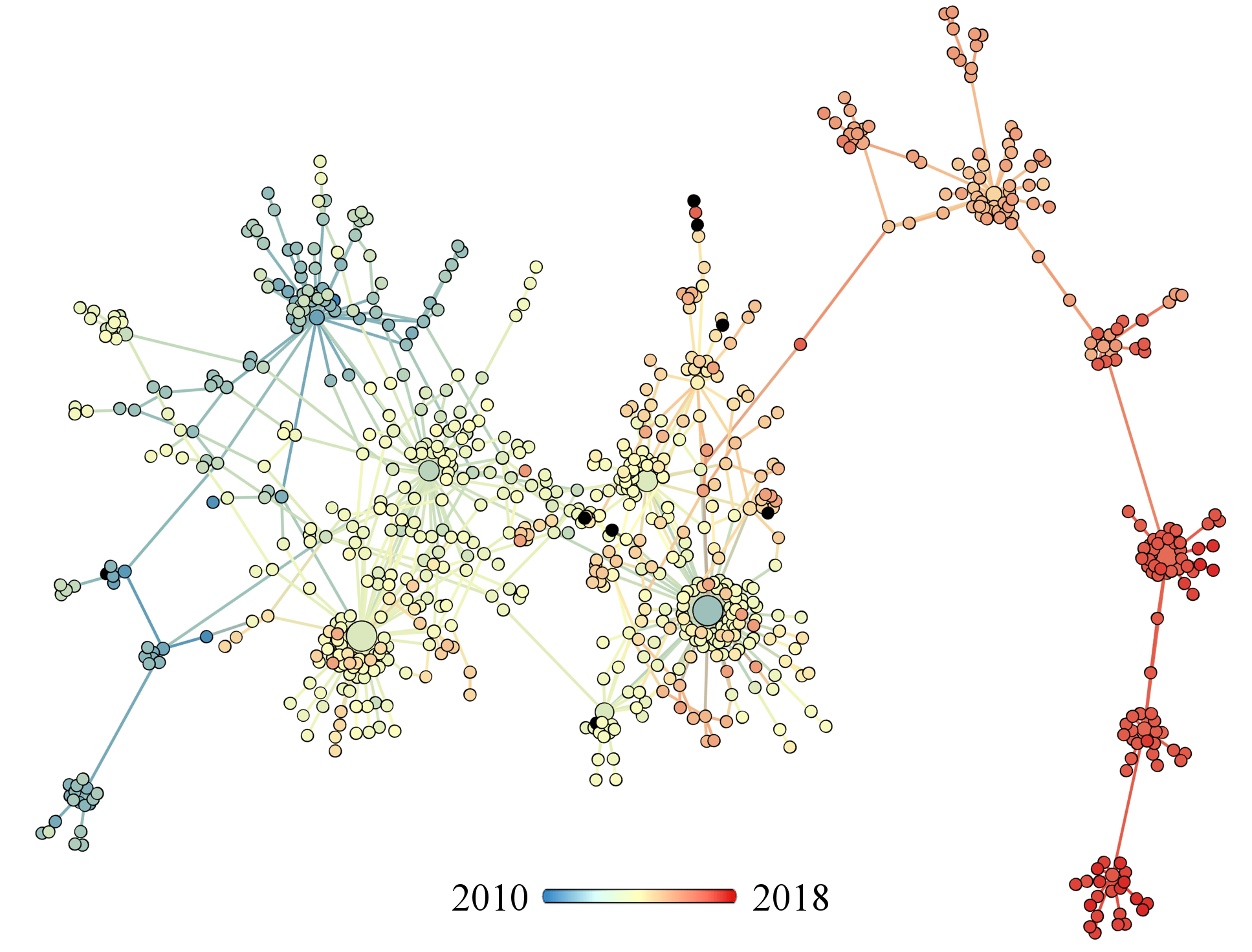}

\includegraphics[width = 0.8\linewidth]{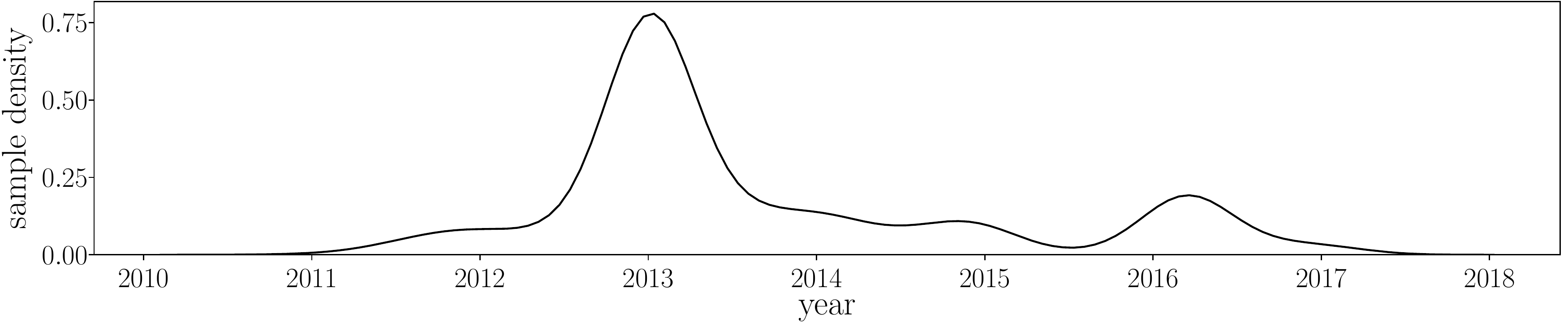}

\caption{\textbf{Sample dates among strains of second largest network component.} \textbf{(top)} Nodes colored by first sample date (8 nodes with lacking sample dates colored black), with a larger radius corresponding to more samples (max sample count 337). \textbf{(bottom)} Sample date distribution across all dated samples of strains within the above network.}
\label{fig:network}
\end{figure}

\subsection{Network topology in time}

The genotype network grows in time as new strains emerge and are sampled. For example, the growth of the second largest component is shown in Fig.~\ref{fig:network}, with each node colored by the first sample date for each strain. This component is large enough to span several years while remaining small enough to qualitatively observe network growth in time. The blue-shifted nodes represent the earliest observed strains among those belonging to this component, the first of which was sampled in late 2010. The majority of unique strains were sampled from 2012 to 2015, including multiple high-degree strains and their neighbors. The most recent strains from this network component are red-shifted, clearly depicting the tree-like growth process. 

Numerous hubs are seen throughout the network, with the largest hubs existing around the 2012-2013 flu season that contributed numerous strains to this component (Fig.~\ref{fig:network}, bottom). Seasonality is reflected in the sample date distribution of this component, with multiple peaks around the start of the calendar year during flu season.

\begin{figure*}[t!]
\centering
\includegraphics[width = 0.9\linewidth]{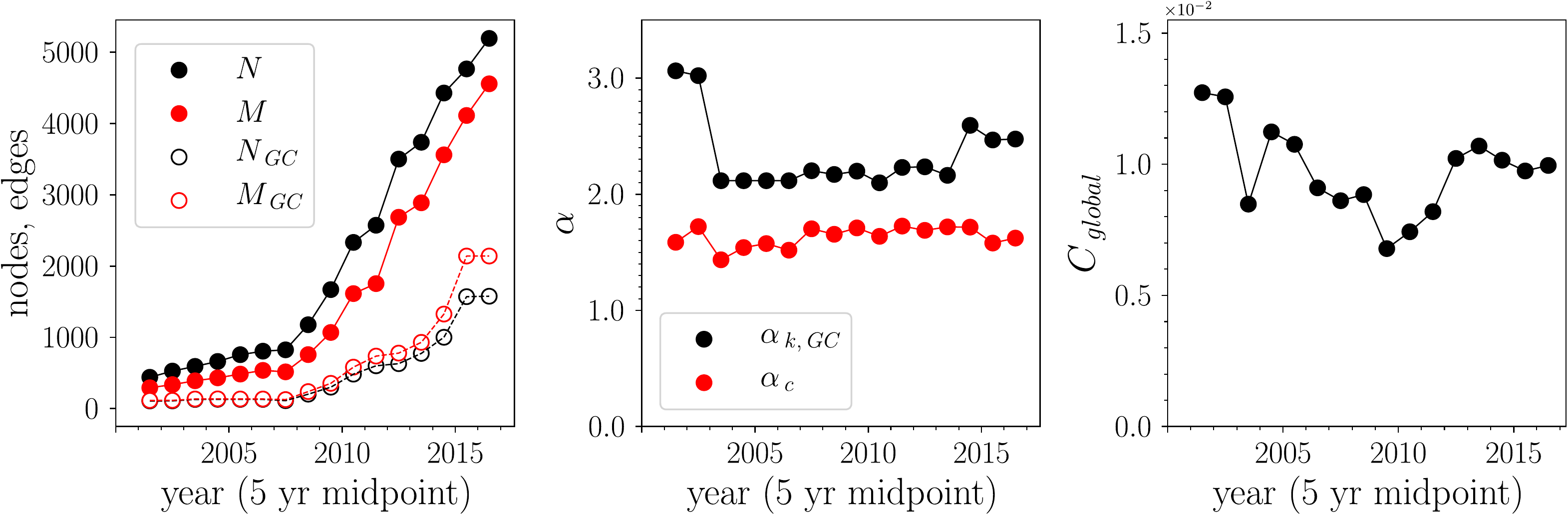}
\caption{\textbf{Network statistics in time.} INFV A (H3N2) HA genotype networks generated using samples within a sweeping 5-year window from July 1999 through June 2019, shown at midpoint. \textbf{(left)} Number of nodes and edges for entire network and giant component. \textbf{(center)} Power-law scale exponents $\alpha_c$ and $\alpha_{k,\; GC}$ obtained by fitting the tail of the distribution above $c_{min}=k_{min}=7$ following Fig.~\ref{fig:features}. \textbf{(right)} Global clustering coefficient $C_{global}$ over time.}
\label{fig:time}
\end{figure*}

Features of the genotype network remain fairly stable in time, even in the presence of a constantly increasing sampling rate. Genotype networks were constructed using samples within a 5-year window, sweeping across the entire sample set. These temporally restricted genotype networks display the structure of the network local in time -- an important consideration given that strains emerge and fall out of circulation. These networks display the increased availability of sequenced samples with each successive year, with notable increases in sampling since 2008 (Fig.~\ref{fig:time}, left panel). The number of nodes and edges has grown steadily in the past 2 decades across both the entire network of the 5 year windows and its giant component. 

Scaling of both degree distribution and component size distribution tails remain fairly constant in time. The scale exponent for degree averaged 2.34 across these networks, varying from $2.10 < \alpha_k < 3.06$. We find over a decade of consistency near its mean (about 2.2 or 2.5) even as the network grew several times larger, Fig.~\ref{fig:time}(center panel). Similarly, the power-law exponent for component size averaged 1.63 and varied within $1.44 < \alpha_c < 1.73$, demonstrating consistency in time and a comparable independence from sample rate as the network grew, Fig.~\ref{fig:time}(center panel). Here $c_{min}$ and $k_{min}$ were fixed at 7, enabling a direct comparison with the entire network. 

Local cycles continue to remain prevalent in the network through time. The global clustering coefficient varied within $6.78\times 10^{-3} < C_{global} < 1.27\times 10^{-2}$, showing greater variability than scaling factors, Fig.~\ref{fig:time}(right panel). Similarly, degree assortativity varied within $-0.365 < r < -0.124$, demonstrating variability but preserving the disassortative structure of the network. The above features demonstrate that in the presence of variable sequence sampling rates, genotype networks possess fairly consistent topological features that are highly predictable from  recent years.

\subsection{Epidemics in random graphs}

To investigate how the genotype network structure may be influenced by the spread of disease and learned host immunity, a multistrain SIRS model was constructed with an underlying network of strains. The incorporation of a genetic strain structure allows for both mutation between neighboring strains and cross-protective immune effects, defined as a function of network distance.

\begin{figure}[t!]
\centering
\includegraphics[width = 0.64\linewidth]{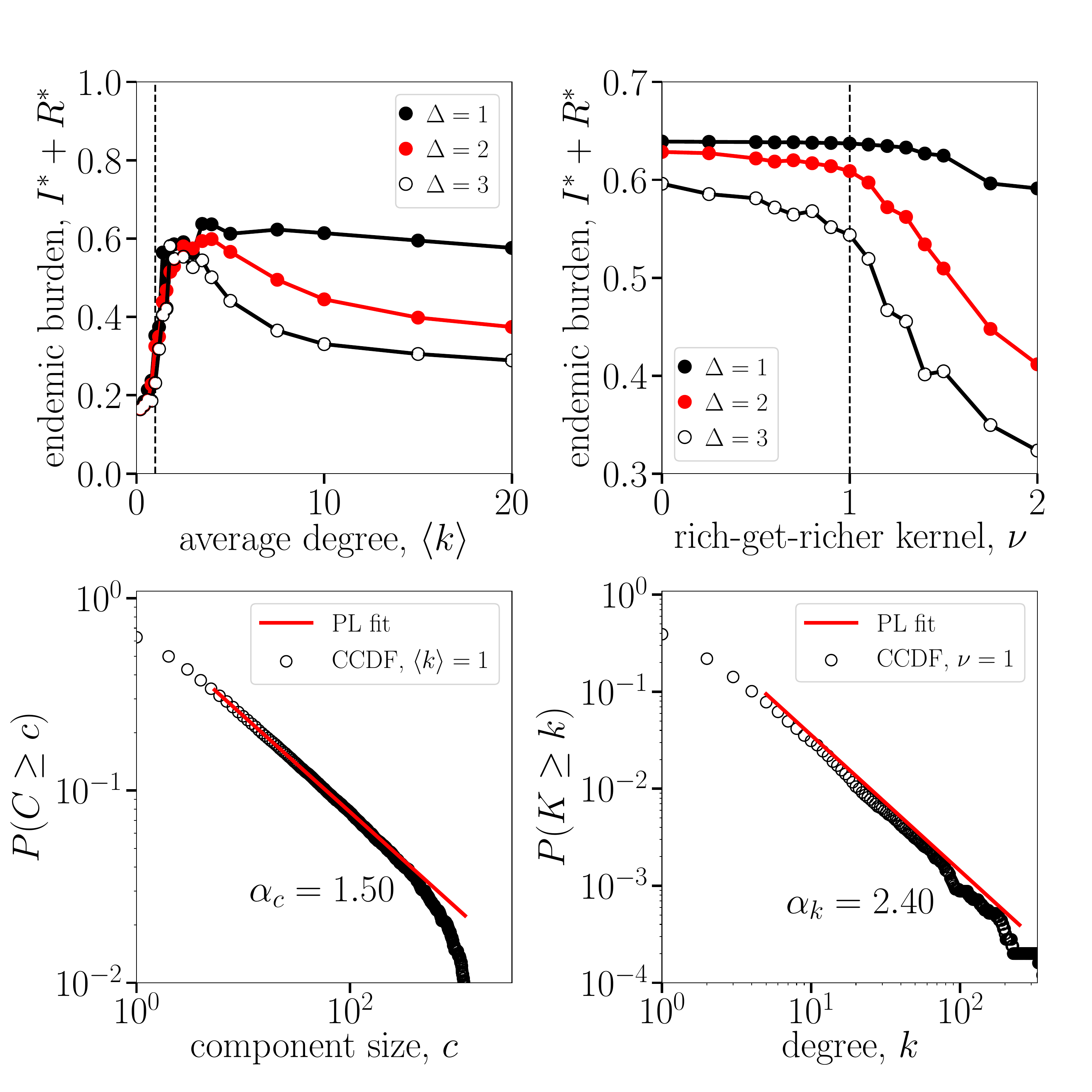}
\caption{\textbf{Endemic disease burden as a function of the connectivity and heterogeneity of the underlying genotype network. }
(\textbf{top left)} We look at the disease burden $I^*+R^*$ in a SIRS model with an underlying, random, genotype network. The network is specified as an Erd\H{o}s-R\'enyi random graph with varying average degree. (\textbf{top right)} The network is now generated by a non-linear preferential attachment scheme with a fixed density (corresponding to the empirical INFV genotype network) and varying attachment kernel $\nu$. In this scheme, $\nu=0$ corresponds to uniform attachment, $\nu=1$ to scale-free networks and $\nu=2$ to star-like networks. Other parameters: Network size $n=250$, mutation rate $\mu=1/50$, transmission rate $\beta=1/2$, recovery rate $\gamma=1/6$, immune loss rate $\alpha=1/100$. (\textbf{bottom left)} Component size of random nodes found in the networks with highest endemic burden in the top left panel, i.e. $\langle k \rangle = 1$. (\textbf{bottom right)} Degree distribution of nodes found in the networks with strongest preferential attachment before endemic burden decreases due to condensation in the genotype network, i.e. $\nu = 1$.
}
\label{fig:connectivity}
\end{figure}

The connectivity or edge density of a genotype network may influence its endemic infection capacity, as suggested by cross-protective immune effects and the observed criticality within the genotype network structure. Here the effects of connectivity were investigated with the implementation of the multistrain model on fully random networks, namely $G(n,p)$ Erd\H{o}s–R\'enyi random graphs~\cite{erdos1960evolution}, with a given number of nodes $n$ and edge probability $p$ controlling connectivity for an average degree of $\langle k \rangle = p(N-1)$. We measure the endemic disease burden $I^* + R^*$, summed over all strains, once the epidemic dynamics has reached an equilibrium (i.e., after a long period of transient dynamics). This disease burden was observed across varying edge densities and levels of immunity transcendence in Fig.~\ref{fig:connectivity}(top left) to determine the relationship between connectivity and endemic infections for a genotype network of a given size.

Non-trivial dynamics are revealed by the multistrain epidemic model with an underlying genotype network structure of Erd\H{o}s–R\'enyi random networks. Endemic disease burdens are lowered in random genotype networks in the presence of high connectivity and non-zero transcending-immunity parameter $\Delta$, producing cross-protective immune effects that outweigh the increase in mutations. On the opposite end, extremely low connectivity also lowers disease burden through increased network fragmentation, resulting in numerous components that restrict mutation pathways between all strains. Together these dynamics produce an optimal connectivity that maximizes disease burden. While slightly affected by parameters, we find that the optimal average degree is observed increasingly close to $\langle k \rangle = 1$ as the pervasiveness of immunity $\Delta$ increases. This density is a critical point of the network structure where a giant component emerges. Around this critical transition, we find a power-law distribution of component sizes with exponent 1.5. This distribution shown in Fig.~\ref{fig:connectivity}(bottom left) is similar to that empirically observed in Fig.~\ref{fig:features}(center).

This critical component size distribution is not found at an average degree $\langle k \rangle = 1$ in the INFV genotype network since its structure is far from that of Erd\H{o}s–R\'enyi random networks. Most notably, the degree distribution of the real network is not homogeneous: The power-law degree distribution shown in Fig.~\ref{fig:features} is radically different from the Poisson degree distributions of Erd\H{o}s–R\'enyi networks. We therefore turn to a non-linear preferential attachment model where networks are grown according to a rich-get-richer process where new strains are a mutation of existing strains chosen randomly but proportionally to their current degree to some power $\nu$, controlling the network heterogeneity~\cite{krapivsky2001degree, krapivsky2001organization}. In Fig.~\ref{fig:connectivity}(top right), we find that the strongest rich-get-richer effect that a genotype network can support before decreases in disease burden is a linear attachment effect, reminiscent of the relationship observed in Fig.~\ref{fig:features}(right). Under this linear preferential attachment, we find a power-law degree distribution with scale exponent 2.43, close to the exponent of 2.3 observed in the INFV genotype network in Fig.~\ref{fig:features}.  

\begin{figure}[ht!]
\centering
\includegraphics[width = 0.5\linewidth]{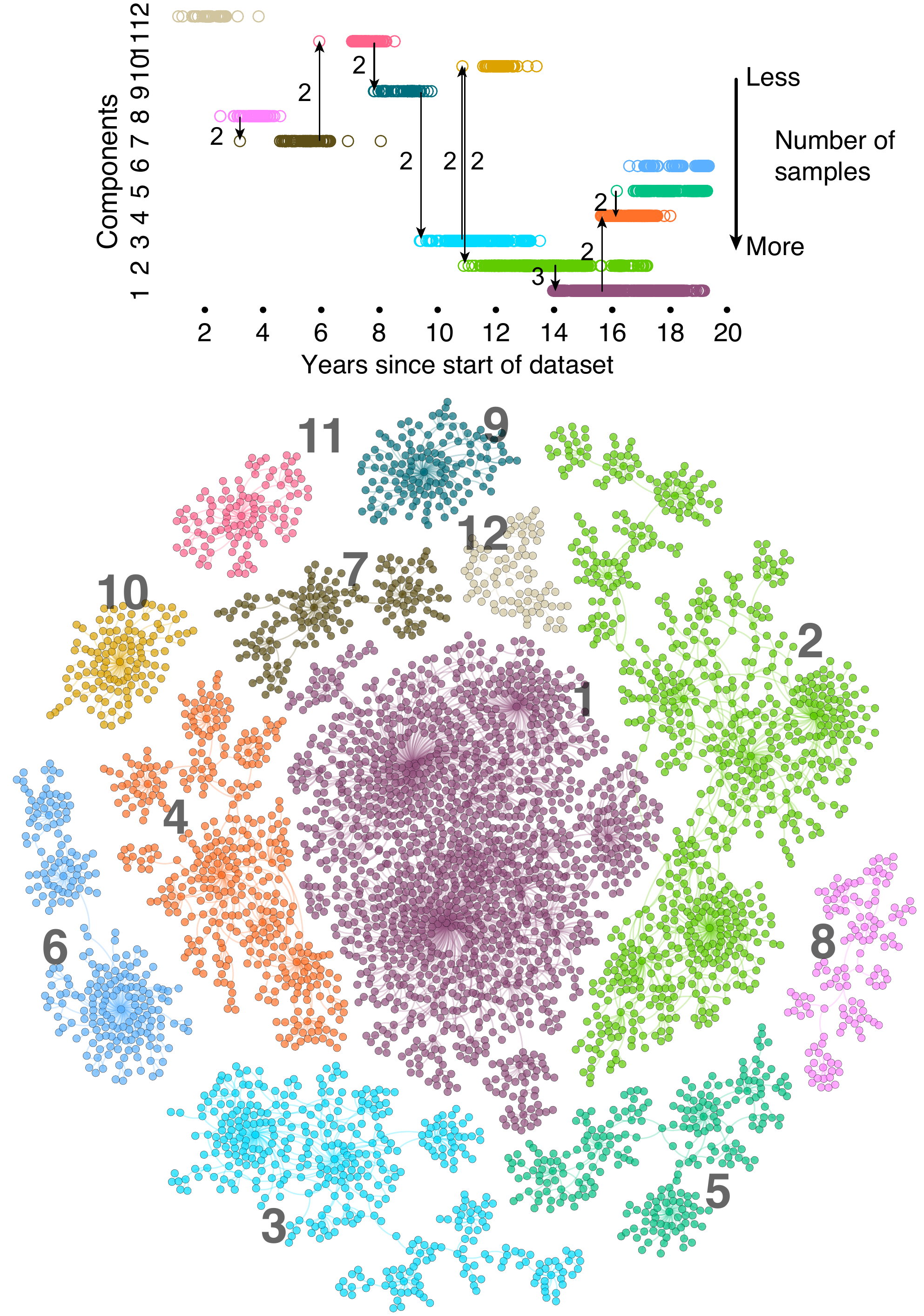}
\caption{\textbf{Higher-order mutations help explain the global structure of the genotype network.}
We show the 12 largest network components. In the top panel, we show how double and triple mutations help explain almost all jumps across components. We also note that components are never re-discovered after more than a few months without new strains emerging therein. Altogether, this analysis suggests that the sampling of strains might be better than originally expected, but also that higher-order networks structures (paths of multiple mutations) might eventually help us better understand the global patterns of INFV immune evasion.
}
\label{fig:flumap}
\end{figure}

\section{Discussion}

In this study, we utilize a large data set and a genotype network to examine INFV evolution from 1999 to 2020. In doing so, we reveal features suggestive of a fundamental structure underlying INFV genotypic space, and by extension, virus evolution.  The INFV genotype networks explore a subspace of all networks that is predictable in structure as they grow in time. Features such as scale-free degree distributions and component size distributions, both suggestive of underlying critical phenomena~\cite{dorogovtsev2008critical}, remained present and consistent in networks generated using temporal subsets of strain samples. 

Given the numerous mutations possible, it may not be realistic to use genotype networks to predict new strains with meaningful accuracy. However, it may be possible to predict their genetic relationship to strains existing in the network structure. Any such efforts would effectively create a map of the genotype space currently occupied by INFV, and suggestive of where in that space it may evolve (see Fig.~\ref{fig:flumap}). Assuming the genetic distance is proportional to antigenic distance~\cite{peeters2017genetic,neher2016prediction, bedford2014integrating}, this is a consequential development with regards to our understanding of cross-protective immune effects and vaccination strain selection. That is, the outlined approaches may offer perspective on which specific genotypes of a given INFV strain might offer the best cross-protective immunity.  


The predictable statistics of the genotype network topology indicates that the INFV A (H3N2) HA genotype network may be influenced by strain-transcending immunity. This is further suggested by the dynamics of a multistrain epidemic model. As the pervasiveness of learned immunity increases, the peak endemic burden expected in our multistrain model shifts closer to, and becomes narrower around, a critical network density corresponding to the emergence of a giant component and a power-law distribution of component sizes. In the future, knowing what mechanisms help shape the genotype network could allow network inference frameworks to identify critical new strains as they emerge \cite{young2019phase, cantwell2021inference}.

The strong positive relationship between degree and sample count implies preferential attachment based on degree, however node age implements a consequential maximum age at which a node may acquire new neighbors. This corresponds to the point at which the strain is not widely circulating or extinct in the host population. Furthermore, the multistrain model suggests that strain-transcending immunity drives this strain extinction process as cross-protective effects increase population immunity towards strains in time. As shown by the model, any stronger preferential attachment mechanisms would also decrease the expected epidemic burden.

This study offers improvements over other methods for studying pathogen evolution. These include phylodynamics, antigenic cartography,~\cite{smith2004mapping, fonville2014antibody} and other network approaches. With regards to phylodynamics, the approach requires few of the population genetic (and other assumptions) that are embedded in phylodynamic approaches. And it offers improvements over existing network models through the additional insights: the identification of power-law properties in INFV genotype networks, and the offering of a mechanism for underlying structure. Our observations suggest that INFV genotype networks explore a subspace predictable in structure, influenced by the effects of strain-transcending immunity. A more realistic network growth process (involving, for example, convergent evolution for the creation of squares) would be necessary to better fit the observed genotype network structure. Likewise, this observed structure is also impacted by the imperfect sampling of INFV strains. Future efforts may utilize more densely sampled populations. 

In summary, we suggest that increased genomic surveillance of multistrain pathogens will allow for similar analyses of other diseases with variable antigenic properties. As the evolutionary forces acting multistrain pathogens differ, we may expect differing network structures from pathogen to pathogen. For instance, HIV has unique pressures from lifetime infection and pathogen evolution, highly active antiretroviral therapy used in its management, as well as bottleneck transmission events and selection biases~\cite{carlson2014selection} -- all mechanisms that could lead to unique network features. Rapidly changing pathogenicity and virulence in emergent viruses, such as SARS-CoV-2, could yield dynamic network features. As the COVID-19 pandemic has generated data at an unprecedented pace and level of granularity, it may offer the opportunity for an analogous comparison~\cite{yin2020genotyping}. 

More broadly, our findings support the importance of multiple methods -- utilizing both existing phylodynamic approaches and network and graph theoretical methods -- towards a comprehensive picture of virus evolutionary dynamics. The use of multiple methods can be complementary, as standard canon from evolutionary theory and methods from complex systems can each offer useful information about pathogen evolution. 

In the future, we might be able to characterize the underlying physics of RNA virus infection networks that can be used to predict long-term patterns, towards improved public health interventions: vaccine strain selection, analysis of evolutionary trajectories, and refinement of the understanding of cross-protective immunity.

\section*{Acknowledgements}
BW and LHD acknowledges support from the National Institutes of Health 1P20 GM125498-01 Centers of Biomedical Research Excellence Award. The authors would like to thank Samuel V. Scarpino for helpful discussions.

\appendix

\section{Statistical methods}
Distribution tails were fitted with power laws using the `poweRlaw' package~\cite{poweRlaw, clauset2009power}. For the aggregated network across all years, we fit power-law tails using after a minimum value of at least 5, found to provide the best goodness of fit at $k_{\textrm{min}}=c_{\textrm{min}}=7$ for both degree and component size. These minimum values were then constrained to $7$ for networks consisting of 5 years of data in our temporal analysis.

\section{Multistrain epidemic model}

The dynamics of the model described in the main text and Ref.~\cite{williams2021localization} can be tracked with the following set of ordinary differential equations,
\begin{align}
    \frac{dS}{dt} &= -\beta \sum_{i=1}^{N} \frac{SI_i}{N} + \alpha \sum_{i=1}^{N}R_i \nonumber\\
    \frac{dI_i}{dt} &= \beta \frac{SI_i}{N} - \gamma I_i + \mu \sum_{j=1}^N A_{i,j}(I_j-I_i) + \sum_{j=1}^N \beta^*_{ij} \frac{I_iR_j}{N} \nonumber\\
    \frac{dR_i}{dt} &= \gamma I_i -\alpha R_i - \sum_{j=1}^N \beta^*_{ij} \frac{I_jR_i}{N} \; ,\nonumber
\end{align}
with $\beta^*_{ij} = \beta \left( 1-e^{-x_{ij}/\Delta} \right)$.

\section{Experiment on Erd\H{o}s-R\'enyi networks}
In the left column of Fig.~\ref{fig:connectivity} we present the endemic disease burden $\sum_i I_i+R_i$ (i.e., recent infections) of our multistrain epidemic model on Erd\H{o}s-R\'enyi networks~\cite{erdos1960evolution}. The endemic state is defined as the fixed point where all derivatives of the system are equal to zero. Erd\H{o}s-R\'enyi networks are obtained by generating a set of $n = 250$ nodes and connecting each possible pair of nodes with probability $p = \langle k \rangle / (N-1)$ such that the expected degree of all nodes (number of first network neighbors) is set by $\langle k \rangle$. 

\section{Experiment on preferential attachment networks}
In the right column of Fig.~\ref{fig:connectivity} we present the endemic disease burden $\sum_i I_i+R_i$ of our multistrain epidemic model on networks grown through preferential attachment~\cite{krapivsky2001degree, krapivsky2001organization}. These networks are obtained by starting from a pair of connected nodes and growing the network until we reach a network of size $n = 250$ nodes.

The networks are grown through the following discrete stochastic process. At each time step, we either connect an existing pair of nodes with probability $p$ or connect a new node to an existing node with complementary probability $1-p$. The probability $p$ sets the expected density of the network, since after $t$ time step we expect $t$ edges and $(1-p)t$ nodes for an average degree $\langle k \rangle = 2/(1-p)$. In our experiment $p$ is chosen to fix the average degree to that observed in the giant component of our empirical data, i.e. $\langle k \rangle = 2.73$. 

At every time step, we therefore need to pick either two existing nodes (probability $p$) or one existing node (complementary probability $1-p$). These existing nodes are chosen proportionally to their degree $k$ proportionally to the kernel $k^\nu$. Meaning a given node $i$ of degree $k_i$ will be chosen with probability $k_i^\nu/\sum_j k_j^\nu$. A kernel with $\nu=0$ correspond to uniform attachment whereas the linear kernel $\nu=1$ correspond to the much studied linear attachment model of Barab\'{a}si and Albert~\cite{BA}.


\end{document}